\documentclass[aps,rmp,twocolumn]{revtex4}
\usepackage{graphicx}
\newcommand\beq{\begin{equation}}
\newcommand\eeq{\end{equation}}
\newcommand\bear{\begin{eqnarray}}
\newcommand\eear{\end{eqnarray}}
\makeatletter       

\renewcommand{\p@subsection}{}

\renewcommand{\p@subsubsection}{}

\makeatother

\begin{document}

\title{Comparative study of the electron conduction in azulene and naphthalene}
\author{{\bf SUDIPTA DUTTA, S. LAKSHMI and SWAPAN K. PATI}\\
Theoretical Sciences Unit, Jawaharlal Nehru Centre For
Advanced Scientific Research,\\
Jakkur Campus, Bangalore 560064, India,\\
DST unit on Nanoscience, \\Jakkur Campus, Bangalore 560064, India}

\widetext

\begin{abstract}
\parbox{6in}

{\bf Abstract. We have studied the feasibility of electron conduction in azulene molecule
and compared with that in its isomer naphthalene. We have used non-equilibrium 
Green's function formalism to measure the current in our systems as a response 
of the external electric field. Parallely we have performed the Gaussian 
calculations with electric field in the same bias window to observe the impact 
of external bias on the wave functions of the systems. We have found that, the 
conduction of azulene is higher than that of naphthalene inspite of its intrinsic 
donor-acceptor property, which leads a system to more insulating state. Due to 
stabilization through charge transfer the azulene system can be fabricated as a 
very effective molecular wire. Our calculations shows the possibility of huge 
device application of azulene in nano-scale instruments.

\bigskip

}

\end{abstract}

\maketitle

\narrowtext

\bigskip

{\bf 1. Introduction}

\bigskip

Electron transport through a single molecule has generated a great deal of interest
both theoretically and experimentally in recent times because of its wide 
variety of device applications\cite{Ratner1994,Alivisatos1998,
Joachim1997,Jortner1997,Joachim1995,Zhou1997,Dorogi1995}. These 
types of molecules are believed to be
the potential components of future nanoscale computational or electronic 
devices. Recent advances in experimental techniques have allowed fabrication
and measurement of current through such nanoscale systems. Various molecules 
have already been demonstrated to behave as wires, switches, diodes, RAMs 
etc\cite{Reed1997,Cui2001,Kushmerick2002}. In most of the experiments, the 
nanoscale material is an organic molecule or a $\pi$-conjugated polymer. Many 
theories too have been developed from empirical to semiempirical and $ab$ $initio$ 
level to describe the electrical response functions of the 
nanomaterials\cite{Lakshmia2005,Lakshmi2004,Lakshmib2005,Sengupta2006,Sudipta-JPCMa,Sudipta-JPCMb,Sudipta-JPCB}.

The ability of a molecule to switch between off and on states is one of the
most important applications of these nanoscale materials. Experimentally, this 
has been observed in many organic molecules with various donor and acceptor 
substituents. Such unusual interesting behaviour of organic molecules with donor and 
acceptor moieties stimulates us to study the electronic conduction in azulene 
molecule and to compare that with its isomer, naphthalene. Although 
naphthalene and azulene are structurally similar with the same number of carbon 
and hydrogen atoms and the same number $(ten)$ of $\pi$ electrons, the 
properties of the azulene molecule differ from that of its isomer in many 
respects\cite{Vogtle1989,Streitwieser1961,Salem1966,Bergman1971}. 
The geometry of the azulene molecule consists of fused five and seven 
membered rings which contributes to the intense blue color and the large 
dipole moment $(\mu=0.8-1.08 D)$ for a hydrocarbon system\cite{Tobler1965,Anderson1959}. 
It has interesting photophysical properties like $S_{2} \rightarrow S_{0}$
fluorescence\cite{Beer1955}, large hyperpolarizability\cite{Morley1989} etc.
The observed 
dipole moment is due to the partial charge transfer from the seven 
membered ring to the five membered ring which gives the molecule aromatic 
stability obeying Huckel $4n+2$ rule in both the rings with $n=1$ and this
gives the molecule intense blue color. 
Therefore, the ground state consists of a donor seven membered ring and
an acceptor five membered ring fused together. On the contrary, naphthalene 
molecule consists of two fused six membered rings with six electrons in each 
ring. Here, there is no occurence of charge transfer to gain aromatic stability
which ensures the zero dipole moment and thus making it appear colorless. 

Due to the interesting charge transfer properties of the two thiolated 
molecules, we have studied their I-V characteristics between two gold 
electrodes using the non-equilibrium Green's function methods, already 
widely reported in the literature. We apply the electric field along the 
principle $C_2$ axis in both the directions. Parallely, we use quantum 
chemical methods to study the effect of electric field on their molecular 
orbitals that are accessed during the transport.

In section II we have discussed about our computational details. Section III and
IV contain the results with discussion and conclusion of our work respectively.


\bigskip

{\bf 2. Computational Method}

\bigskip

\begin{figure}
\centering
\includegraphics[scale=0.2]{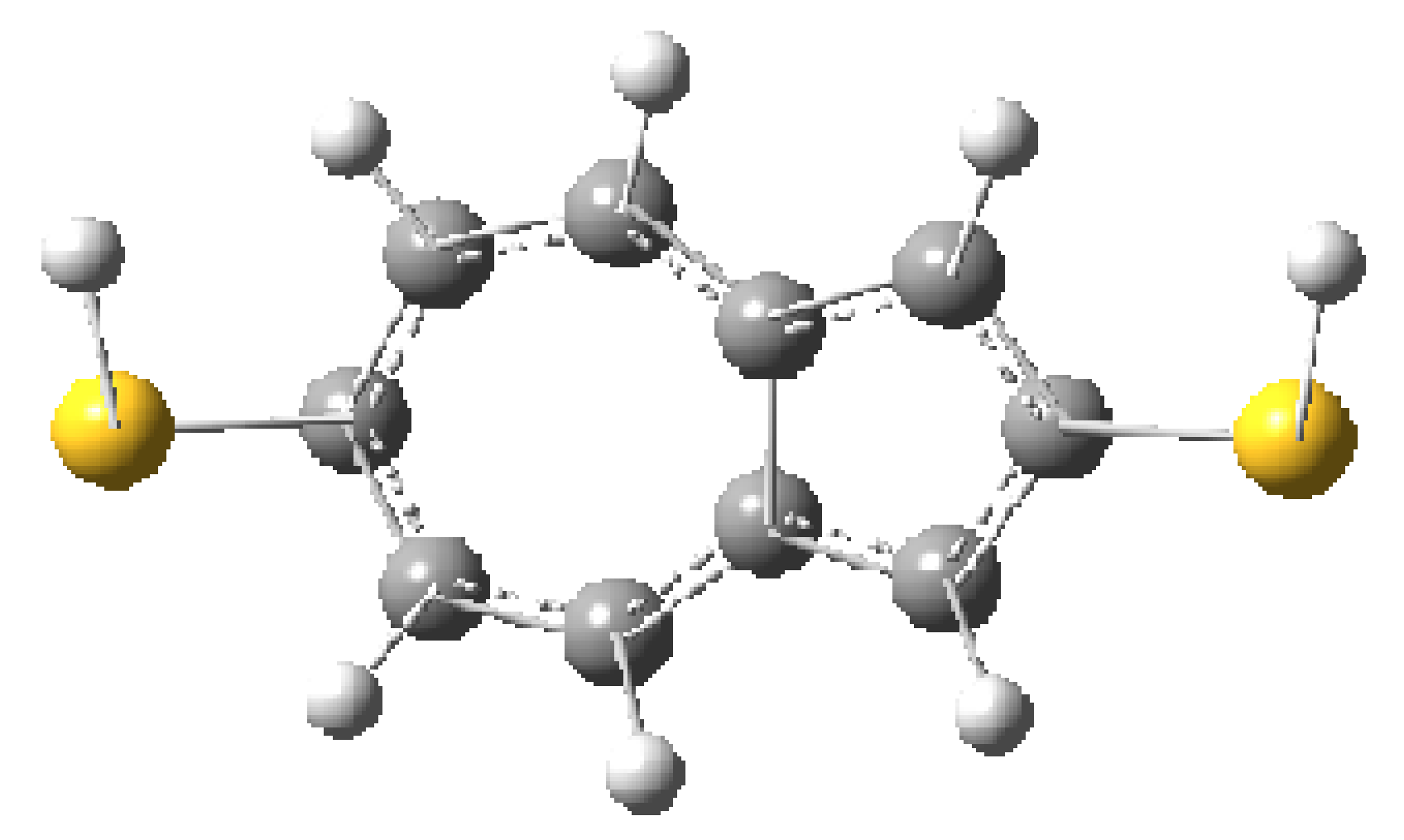}
\includegraphics[scale=0.2]{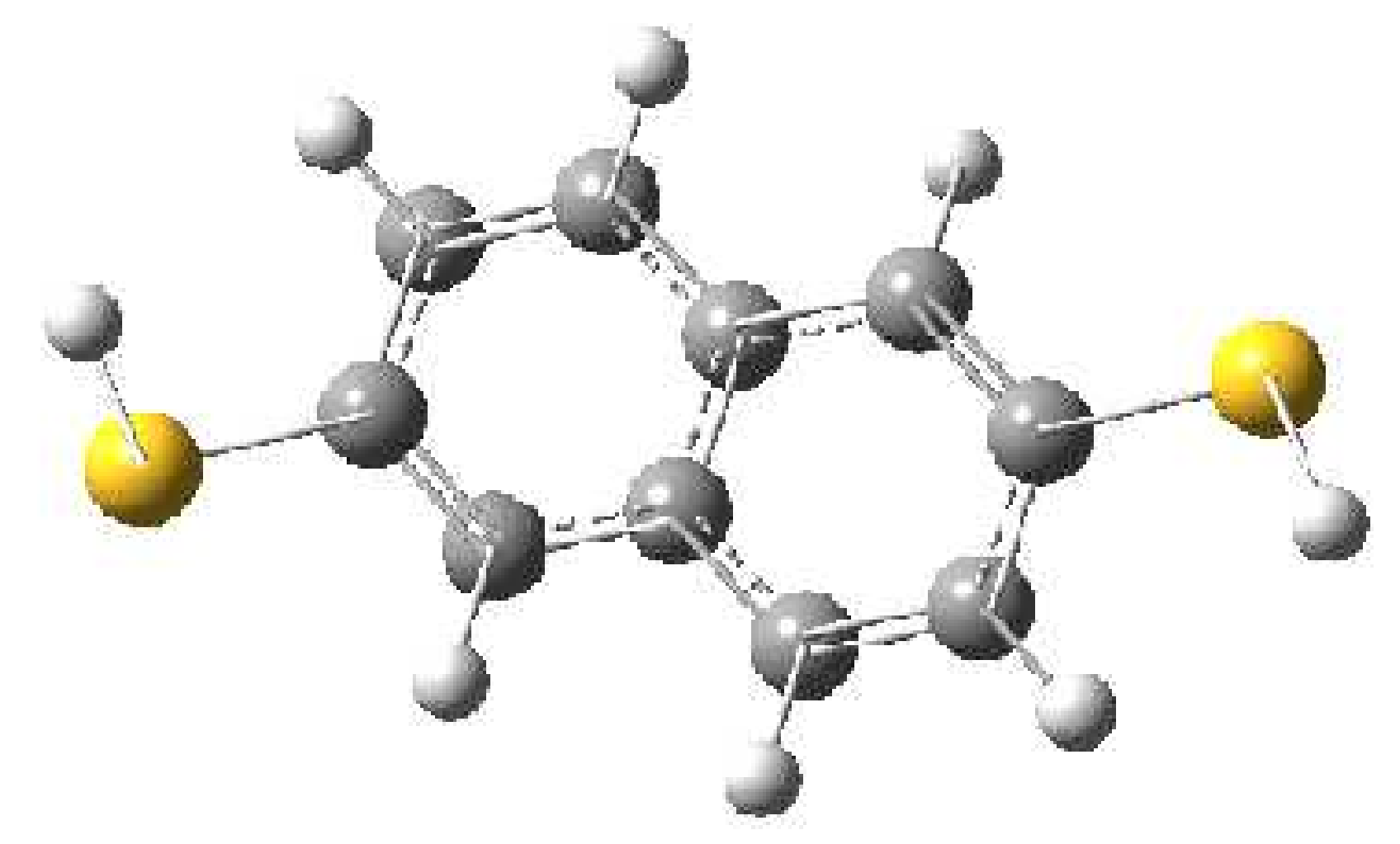}
\caption{Structure of azulene and naphthalene with two thiol groups 
linked at furthest positions}
\end{figure}

Firstly, to study the effect of electric field on the frontier molecular 
orbitals of both the molecules, we optimized the ground state structures 
of azulene and naphthalene with two thiol groups placed at the furthest 
positions as shown in fig.1 using the density functional methods with
6-31++G(d) basis set and density functional exchange correlation B3LYP. 
We then optimize the structures of both the molecules using the same 
basis set in the presence of electric field, applied along the chain axis 
of the molecules. We vary the electric field in the interval of $0.001 a.u.$
and $1 a.u.= 5.142 X 10^{11} Volt/m$. Vibrational analysis is then performed 
on each optimized structure to check whether, what we obtain are the stable 
minimum energy structures with all real frequencies. Then we calculate the 
Fock and overlap matrices for the ground state molecules which we use later 
for the calculation of current using non-equilibrium Green's function $(NEGF)$ 
method\cite{Supriyo1995}. All calculations are performed with the standard 
Gaussian-03\cite{Gaussian2003} package.

Parallely, we study the equilibrium properties of the molecules before 
connecting them with the electrodes. We first calculate the Green's function 
$G$ in equilibrium situation using the formula,

\begin{eqnarray}
G &=& \frac{1}{ES - H + i\eta}
\end{eqnarray}

\noindent where, $E$ is the electron injection energy, $S$ and $H$ are the 
overlap and Fock matrices respectively. $\eta$ is a small arbitrary broadening 
parameter arises mostly due to temperature. We get obtain $S$ and $H$ from the 
density functional calculations with Gaussian-03 package. The hydrogens attached 
to the thiol groups are removed by modifying the Hamiltonian and overlap matrices
at those positions and these are then used for the calculation of the Green's 
function. From the eigen energies of 
the relaxed system, the Fermi energy, $E_f$ of the electrodes is calculated 
assuming that it falls midway between the equilibrium highest-occupied molecular 
orbital (HOMO) and lowest-unoccupied molecular orbital (LUMO) energies of the 
molecule alone. The Fermi energy can be anywhere between the HOMO and LUMO and 
is generally considered to be a fitting parameter. Then we calculate the density 
of states and the transmission $(T(E))$ between the terminal sulfur atoms ($S_1$ and 
$S_2$).

\begin{eqnarray}
T(E) \propto |G_{S_{1}S_{2}}|^{2} \eta^{2}
\end{eqnarray}

The external electric field 
applied on the system has the form of a ramp potential, distributed over all the
sites in such a way that the potential $V_{i}$ at site $i$ becomes
$-\frac{V}{2}+V\frac{x(i)-x(S_1)}{x(S_2)-x(S_1)}$, where V is the applied voltage 
and $x(i)$, $x(S_1)$ and $x(S_2)$ are the x-coordinates of $i$-th atom, 1st and 
2nd sulfur atoms respectively. This form of the potential
ensures that the bias varies from $-V/2$ to $V/2$ across the molecule.
The potential adds an extra diagonal term $\sum\limits_{i}V_{i}a^\dagger{_i}a{_i}$
to the Fock matrix. 

The effect of electrodes on the molecule has been incorporated 
through the well known self-energy term $\Sigma$ for gold electrodes. With this 
imaginary self-energy terms the Hamiltonian gets modified as,

\begin{eqnarray}
H &=& H_i + \Sigma_L + \Sigma_R
\end{eqnarray}

\noindent where $H_i$ is the initial Hamiltonian for only the molecule. $\Sigma_L$ and 
$\Sigma_R$ are self-energies of left and right electrodes 
respectively. Having obtained the modified Hamiltonian, the current through the 
molecule is obtained using Landauer's formula\cite{Landauer1957},

\begin{eqnarray}
I(V) &=& \frac{2e}{h}\int\limits_{-\infty}^{\infty} dE[Tr(\Gamma_{L}G\Gamma_{R}G^{\dagger})]
\end{eqnarray}

\noindent where $\Gamma_{L}$ and $\Gamma_{R}$ are the anti-Hermitian parts of the 
self-energy matrices, $\Sigma_L$  and $\Sigma_R$ respectively,

\begin{eqnarray}
\Gamma_{L,R} &=& i(\Sigma_{L,R} - \Sigma_{L,R}^{\dagger})
\end{eqnarray}

\noindent which describe  the broadening of the energy levels due to the coupling of the 
molecule to the electrodes. The Green's function is obtained as the inverse of 
the modified Hamiltonian,

\begin{eqnarray}
G &=& \frac{1}{ES - [H + i(\Sigma_L + \Sigma_R)]}
\end{eqnarray}

Using this Green's function we calculate the current.

We calculate the transmission using the formula,

\begin{eqnarray}
T(E) &=& Tr(\Gamma_{L}G\Gamma_{R}G^{\dagger})
\end{eqnarray}

\begin{figure*}
\centering
\includegraphics[scale=0.7,angle=270]{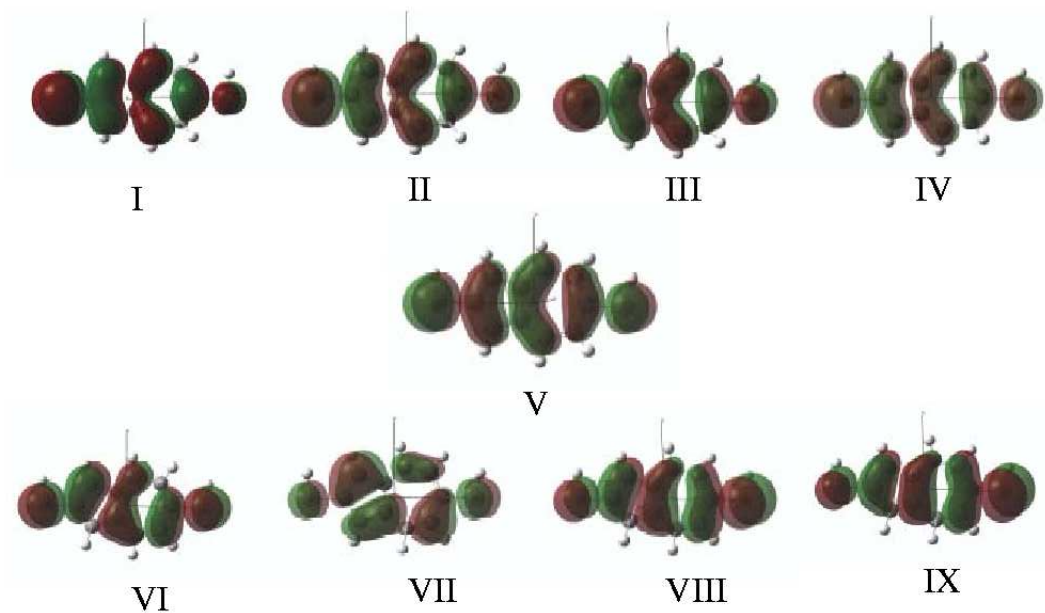}
\caption{HOMO plots for the electric fields 0.01 (I), 0.008 (II), 0.004 (III),
0.002 (IV), 0 (V), -0.002 (VI), -0.004 (VII), -0.008 (VIII), -0.01 (IX) for the azulene
molecule with two terminal thiol groups (unit of electric field is a.u).}
\end{figure*}

\begin{figure*}
\centering
\includegraphics[scale=0.7,angle=270]{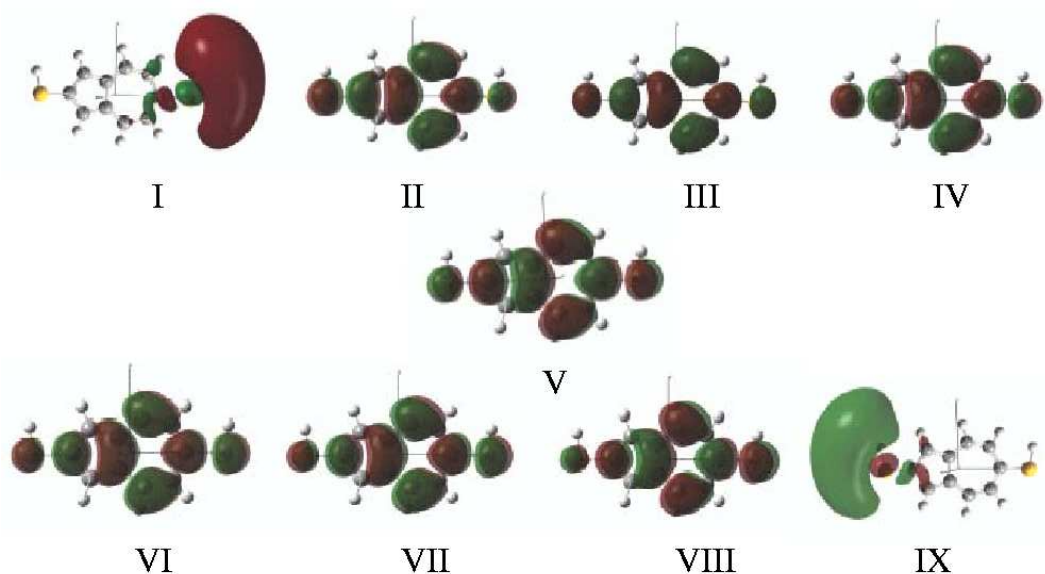}
\caption{LUMO plots for the electric fields 0.01 (I), 0.008 (II), 0.004 (III),
0.002 (IV), 0 (V), -0.002 (VI), -0.004 (VII), -0.008 (VIII), -0.01 (IX) for the azulene
molecule with two terminal thiol groups (unit of electric field is a.u).}
\end{figure*}


\bigskip

{\bf 3. Results and Discussions}

\bigskip

On application of electric field, the HOMO and LUMO of azulene changes
from the ground state as can be seen from fig.2 and fig.3 respectively. 
These figures clearly show that, the effect of electric field, applied 
along positive x-axis differs a lot from that along opposite direction.
In case of HOMO, the electron density is delocalized over the whole molecule 
and it shows electron density shift along the bias direction. LUMO electron 
density is also delocalized but at higher bias value it shows complete 
localization of electron density at one end which depends on the bias direction.
The ground state dipole moment also changes differently for the electric field 
in either directions. From the fig.4 it can be seen that, for both the electric 
field directions, the dipole moment of azulene system
increases. In absence of electric field the azulene molecule has some dipole
moment which increases with increasing bias in positive x-axis direction. But 
the electric field along negative x-axis first reduces the dipole moment followed 
by a monotonic increase which can be seen clearly from the inset of fig.4. This 
observation is in agreement with the fact that the 
electron flows from the seven membered ring to the five membered ring. The electric 
field in the opposite direction forces the electrons to go from five membered ring 
to seven membered ring and thus the dipole moment reduces from the ground state 
dipole moment initially for lower bias. Further increase in bias in this direction 
accumulates the charge on the seven membered ring resulting in a dipole moment 
increase. But the dipole moment of naphthalene molecule increases with increasing 
bias in both the directions from the ground state zero dipole moment. 

\begin{figure}
\centering
\includegraphics[scale=0.3]{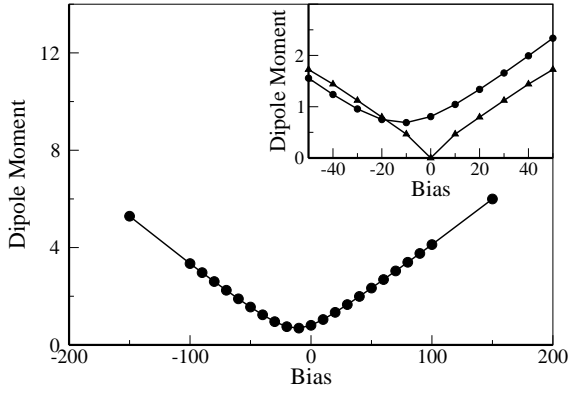}
\caption{Variation of dipole moment of the azulene system as a function of electric
field. Inset shows the dipole moment for both azulene $(circle)$ and naphthalene 
$(triangle)$ in a smaller bias window.}
\end{figure}

Fig.5 shows the variation of the energy of azulene system as a function of 
external electric field. The energy decreases with increasing 
electric field in $+ve$ x-axis, since the system gets stabilization for 
the charge transfer on application of electric field in this direction. But field 
along $-ve$ x-axis initially destabilizes the molecule at smaller bias values followed 
by a monotonic decrease in energy with increasing bias. The initial destabilization 
is attributed to the fact that, the small bias tries to force the electrons to flow 
in its direction in which the system loses its aromatic stability. But for larger 
bias values, the system again behaves like a charge transfer compound and starts 
getting stabilized. Inset of fig.5 shows the comparison of the energy variation of 
both azulene and naphthalene systems as a function of bias. The energy of naphthalene 
molecule decreases from the ground state energy on application of external bias, 
applied in both the directions. 

\begin{figure}
\centering
\includegraphics[scale=0.3]{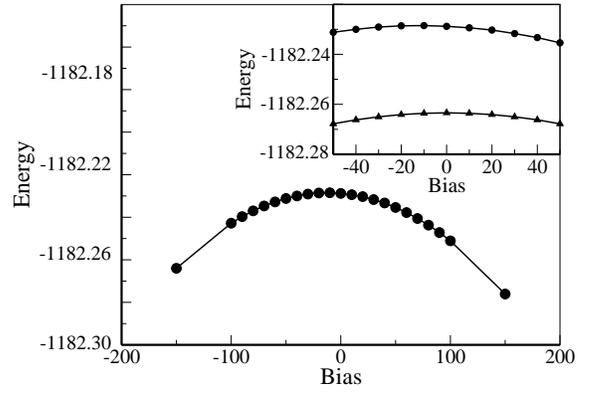}
\caption{Energy of the azulene system as a function of external electric field.
Inset shows the energy for both azulene $(circle)$ and naphthalene $(triangle)$
in a smaller bias window.}
\end{figure}

\begin{figure}
\centering
\includegraphics[scale=0.4]{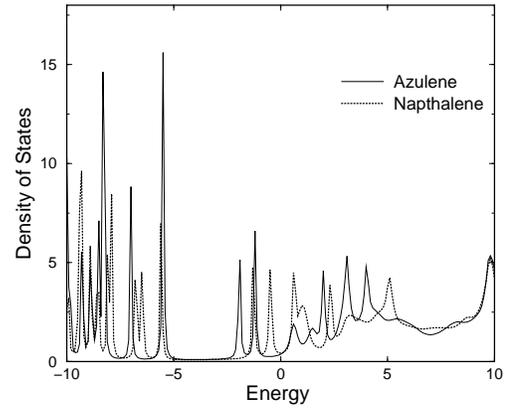}
\caption{Density of states as a function of energy for azulene and naphthalene
molecules with two terminal thiol group.}
\end{figure}

So, it is clear from the above discussion that, the electron density rearranges 
differently for electric fields along $-ve$ and $+ve$ x-axis in case of azulene
system, whereas in case of naphthalene system the effects of electric field along
both the directions are same due to its symmetric structure. Now, we calculate 
the current as a response of applied electric field using the $NEGF$ 
formalism for both azulene and naphthalene systems. First we plot the equilibrium, 
i.e, the zero bias density of states $(DOS)$ in fig.6 and transmission in fig.7.
From the DOS plot it can be clearly seen that the HOMO-LUMO gap 
in case of naphthalene $(\sim 4 eV)$ is higher than that of azulene $(\sim 3 eV)$ 
system. HOMO energy for naphthalene is lower than that of azulene and as a result, 
naphthalene is stabler than its isomer. From the plot of equilibrium transmission, it
is very evident that, the HOMO of both azulene and naphthalene conduct well, whereas 
the LUMO of both show negligible transmission. Hence in our current calculations,
we place the Fermi energy closer to the HOMO to probe conduction through the 
occupied levels.

\begin{figure}
\centering
\includegraphics[scale=0.4]{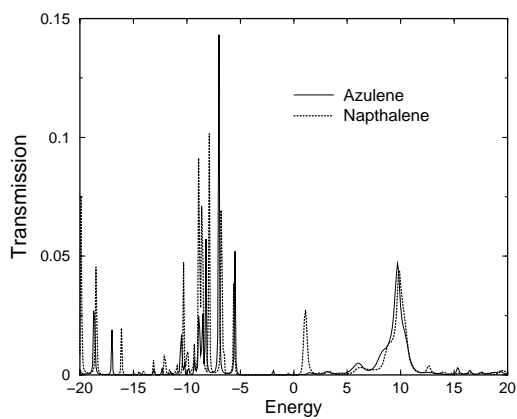}
\caption{Transmission as a function of energy for azulene and naphthalene
molecules with two terminal thiol group.}
\end{figure}

\begin{figure}
\centering
\includegraphics[scale=0.4]{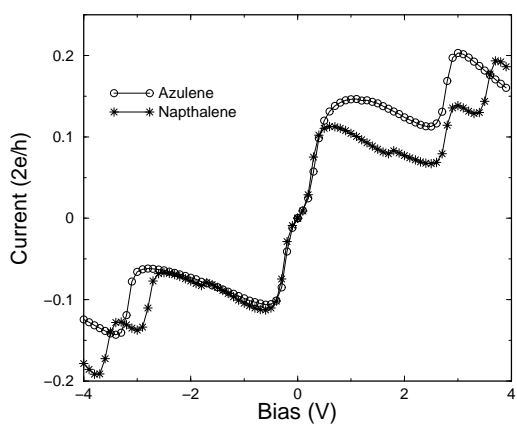}
\caption{Current as a function of bias for azulene and naphthalene molecules
with two terminal thiol groups.}
\end{figure}

To get proper molecular understanding, we keep the bias window more or less same 
for both types of calculations ($NEGF$ and geometry optimization with electric field). 
In our calculations, the length scale 
of the molecules are $8-9 A^{o}$, which ensures the bias window of Gaussian 
calculations similar to the $NEGF$ calculations. Fig.8 shows the I-V 
characteristics with steps for both azulene and naphthalene systems. The I-V
characteristics of naphthalene is completely symmetric in nature for both the bias
directions, whereas for azulene it is not symmetric.
It is very evident that, both the molecules show step-like behavior in I-V
characteristics. These are the well known eigen-value staircases that appear
whenever the electrochemical potential of the electrodes come
in resonance with the molecular levels\cite{Ratner1994}. 
We find that, in lower bias regime, both of them show 
nearly same current. Beyond a certain bias value along positive x-axis the azulene 
system shows more current than the naphthalene system. But for reverse bias the
current of azulene system is lower than the naphthalene. It is due to the fact that,
bias along $+ve$ direction drives the electrons easily from seven membered ring to 
the five membered ring, whereas reverse bias drives the electrons opposite to 
the dipole moment direction. As a consequence, the current becomes lower than 
that of naphthalene which has symmetric electron cloud distribution over the 
two six-membered rings. At higher bias values the current shows a considerable 
decrease with increase in bias due to voltage induced localization. So, the 
interesting result, coming out of our calculations is that, the azulene molecule
which has intrinsic donor-acceptor property, shows more current than naphthalene;
though it is expected to show more insulating behavior. 


\bigskip

{\bf 4. Conclusion}

\bigskip

We have performed both NEGF and Gaussian calculations for both azulene and 
naphthalene systems to measure the current passing through these syatems on
application of external electric field. We have observed the electron density 
rearrangement in both the systems in presence of bias. 
In case of azulene, the electrons flow from seven membered ring to the five 
membered ring more easily resulting in more stabilization with increase in 
bias. But for reverse bias, initially the dipole moment decreases with destabilization 
of the system due to electron flow in opposite direction.
Our $NEGF$ calculations show higher conductivity in azulene than that of in 
naphthalene system. Our results reveal the possibility of potential device application 
of azulene as molecular wire in future nano-scale instruments.

\bigskip

{\bf Acknowledgements}

\bigskip

SKP thanks DST, the Government of India for funding and SD acknowledges the CSIR, the 
Government of India for research fellowship.


\begin{thebibliography}{50}

\bibitem{Ratner1994} Ratner M A $et$ $al.$ 1994 {\it J. Chem. Phys.} {\bf 101} 5173
;Ratner M A $et$ $al.$ 1994 {\it J. Chem. Phys.} {\bf 101} 6849
;Ratner M A $et$ $al.$ 1994 {\it J. Chem. Phys.} {\bf 101} 6856
;Ratner M A $et$ $al.$ 1996 {\it J. Chem. Phys.} {\bf 104} 7296
\bibitem{Alivisatos1998} Alivisatos A P $et$ $al.$ 1998 {\it Adv. Mater. (Weinheim, Ger.)} 
 {\bf 10} 1297; Zahid F, Paulsson M and Datta S 2003 {\it Advanced 
 Semiconductors and Organic Nano-Techniques, edited by H. Morkoc} 
 (Academic Press, New York)
\bibitem{Joachim1997} Joachim C and Roth S 1997 {\it Atomic and Molecular Wires}
 (Kluwer Academic, Dordrecht)
\bibitem{Jortner1997} Jortner J and Ratner M A  1997 {\it Molecular Electronics} 
 (Blackwell, London) 
\bibitem{Joachim1995} Joachim C, Gimzewski J K, Schlitter R R and Chavy C 1995
{\it Phys. Rev. Lett.} {\bf 74} 2102 
\bibitem{Zhou1997} Zhou C, Deshpande M R and Reed M A 1997 {\it Appl. Phys. Lett.} 
 {\bf 71} 611 
\bibitem{Dorogi1995} Dorogi, Gomez J, Andres R P and Reifenberger R 1995 {\it Phys. Rev.}  
 {\bf B52} 9071 
\bibitem{Reed1997} Reed M A $et$ $al.$ 1997 {\it Science} {\bf 278} 252;
Reed M A $et$ $al.$ 2001 {\it Appl. Phys. Lett.} {\bf 78} 3735 
\bibitem{Cui2001} Cui X D $et$ $al.$ 2001 {\it Science} {\bf 294} 571 
\bibitem{Kushmerick2002} Kushmerick J G $et$ $al.$ 2002 {\it Phys. Rev. Lett.} {\bf 89} 086802
\bibitem{Lakshmia2005} Lakshmi S and Pati S K 2005 {\it Phys. Rev.} {\bf B72} 193410
\bibitem{Lakshmi2004} Lakshmi S and Pati S K 2004 {\it J. Chem. Phys.} {\bf 121} 11998
\bibitem{Lakshmib2005} Lakshmi S, Datta A and Pati S K 2005 {\it Phys. Rev.} {\bf B72}, 
045131 
\bibitem{Sengupta2006} Sengupta S, Lakshmi S and Pati S K 2006 {\it J. Phys.: Condens. Mat.} 
{\bf 18} 9189
\bibitem{Sudipta-JPCMa} Dutta S, Lakshmi S and Pati S K 2007 {\it J. Phys.: Condens. Mat.}
{\bf 19} 322201
\bibitem{Sudipta-JPCMb} Dutta S and Pati S K 2008 {\it J. Phys.: Condens. Mat.}
{\bf 20} 075226
\bibitem{Sudipta-JPCB} Dutta S and Pati S K 2008 {\it J. Phys. Chem. B} {\bf 112} 1333
\bibitem{Vogtle1989} Vogtle F 1989 {\it Reizvolle Molekule der organischen Chemie} 
(Teubner, Stuttgart)
\bibitem{Streitwieser1961} Streitwieser A Jr. 1961 {\it Molecular orbital theory for organic 
chemists} (Wiley, New York)
\bibitem{Salem1966} Salem L 1966 {\it The molecular orbital theory for conjugated systems} 
(Benjamin, New York)
\bibitem{Bergman1971} Bergman E D and Pullman B 1971 {\it Aromaticity, pseudo-aromaticity, 
anti-aromaticity} (Academic Press, New York)
\bibitem{Tobler1965} Tobler H J, Bauder A and Gunthard Hs. H 1965 {\it J. Mol. Spectry.}
{\bf 18} 239 
\bibitem{Anderson1959} Anderson A G and Steckler B M 1959 {\it J. Am. Chem. Soc.} {\bf 81}
4941 
\bibitem{Beer1955} Beer M B and Longuet-Higgins 1955 {\it J. Chem. Phys.} {\bf 23} 1390
\bibitem{Morley1989} Morley J O 1989 {\it J. Chem. Soc. Perkin Trans. II} {\bf 103}
\bibitem{Supriyo1995} Supriyo Datta 1995 {\it Electronic Transport in Mesoscopic Systems} 
(Cambridge Studies in Semiconductor Physics and Microelectronic Engineering)
\bibitem{Gaussian2003} Gaussian 03, Revision {\bf B.05},
 M. J. Frisch, G. W. Trucks, H. B. Schlegel, G. E. Scuseria,
 M. A. Robb, J. R. Cheeseman, J. A. Montgomery, Jr., T. Vreven,
 K. N. Kudin, J. C. Burant, J. M. Millam, S. S. Iyengar, J. Tomasi,
 V. Barone, B. Mennucci, M. Cossi, G. Scalmani, N. Rega,
 G. A. Petersson, H. Nakatsuji, M. Hada, M. Ehara, K. Toyota,
 R. Fukuda, J. Hasegawa, M. Ishida, T. Nakajima, Y. Honda, O. Kitao,
 H. Nakai, M. Klene, X. Li, J. E. Knox, H. P. Hratchian, J. B. Cross,
 C. Adamo, J. Jaramillo, R. Gomperts, R. E. Stratmann, O. Yazyev,
 A. J. Austin, R. Cammi, C. Pomelli, J. W. Ochterski, P. Y. Ayala,
 K. Morokuma, G. A. Voth, P. Salvador, J. J. Dannenberg,
 V. G. Zakrzewski, S. Dapprich, A. D. Daniels, M. C. Strain,
 O. Farkas, D. K. Malick, A. D. Rabuck, K. Raghavachari,
 J. B. Foresman, J. V. Ortiz, Q. Cui, A. G. Baboul, S. Clifford,
 J. Cioslowski, B. B. Stefanov, G. Liu, A. Liashenko, P. Piskorz,
 I. Komaromi, R. L. Martin, D. J. Fox, T. Keith, M. A. Al-Laham,
 C. Y. Peng, A. Nanayakkara, M. Challacombe, P. M. W. Gill,
 B. Johnson, W. Chen, M. W. Wong, C. Gonzalez, and J. A. Pople,
 Gaussian, Inc., Pittsburgh PA, 2003.
\bibitem{Landauer1957} Landauer R 1957 {\it IBM J. Res. Dev.} {\bf 1} 223
; Landauer R 1981 {\it Phys. Lett.} {\bf A85} 91

\end{thebibliography}
\end{document}